\documentclass[aps,pre,reprint,groupedaddress,amssymb,amsmath]{revtex4-1}

\usepackage{graphicx}
\usepackage{textcomp}
\usepackage{color}

\begin{document}
 
\title[MD simulations of bubble nucleation]{Molecular dynamics simulations of bubble nucleation in dark matter detectors}

\author{Philipp Denzel}
\email{phdenzel@physik.uzh.ch}
 \author{J\"urg Diemand}
 \author{Raymond Ang\'elil}
\affiliation{Institute for Computational Science, University of Zurich, 8057 Zurich, Switzerland}

\date{\today}


\begin{abstract}
Bubble chambers and droplet detectors used in dosimetry and dark matter particle search experiments use a superheated metastable liquid in which nuclear recoils trigger bubble nucleation.
This process is described by the classical heat spike model of F. Seitz [Phys. Fluids (1958-1988) 1, 2 (1958)], which uses classical nucleation theory to estimate the amount and the localization of the deposited energy required for bubble formation.
Here we report on direct molecular dynamics simulations of heat-spike-induced bubble formation. They allow us to test the nanoscale process described in the classical heat spike model.
40 simulations were performed, each containing about 20 million atoms, which interact by a truncated force-shifted Lennard-Jones potential.
We find that the energy per length unit needed for bubble nucleation agrees quite well with theoretical predictions, but the allowed spike length and the required total energy are about twice as large as predicted.
This could be explained by the rapid energy diffusion measured in the simulation: contrary to the assumption in the classical model, we observe significantly faster heat diffusion than the bubble formation time scale.
Finally we examine $\alpha$-particle tracks, which are much longer than those of neutrons and potential dark matter particles.
Empirically, $\alpha$ events were recently found to result in louder acoustic signals than neutron events. This distinction is crucial for the background rejection in dark matter searches.
We show that a large number of individual bubbles can form along an $\alpha$ track, which explains the observed larger acoustic amplitudes.
\end{abstract}


\maketitle

%
\begin{section}{Introduction}
Liquids can be heated to a temperature higher than their boiling point, without yet going through a phase transition.
The liquids then exist in a metastable, or superheated state, which can be locally disrupted to induce bubble nucleation.
For example, the scattering of cosmic radiation off the liquid nuclei is an energy deposition, which can lead to bubble nucleation.

In the 1950s, this process motivated the use of such superheated liquids to detect particles in bubble chambers \cite{Glaser2}, the most notable being Gargamelle at CERN, which is credited with the discovery of weak neutral currents \cite{Gargamelle2}.
While these detectors have been used extensively in the past, their applications today are limited mostly to monitoring and dosimetry \cite{Apfel}.

However, recent research projects, such as SIMPLE \cite{SIMPLE,SIMPLE2,SIMPLE3}, PICASSO \cite{PICASSO,PICASSO2,PICASSO3}, COUPP \cite{COUPP,COUPP2}, and PICO \cite{PICO} employ bubble chambers (or superheated droplet detectors\textemdash modifications of bubble chambers) in the hope of directly detecting dark matter in the form of weakly interacting massive particles (WIMPs).
While detection in bubble chambers used to consist of event picture taking, nowadays this is achieved by taking advantage of the explosive nature of the phase transition.
The explosions are accompanied by an acoustic shock wave, which is recorded by piezoelectric transducers, located on the chamber walls \cite{piezo}.
Superheated liquid detectors are competitive and complementary to other dark matter detection methods \cite{XENON2,CDMS2,CRESST2}.
However there is still some discussion on several issues, one being the discrimination of WIMP signals from the $\alpha$-particle background \cite{alpha_discr}.

Thanks to increasing computing power, and highly scalable and efficient molecular dynamics codes, it is now possible to simulate phase transitions directly, within a realistic environment\cite{Diemand13_Angelil13, Holyst10, Holyst12}.
Homogeneous liquid-to-vapor nucleation has recently been simulated by some of us \cite{Diemand14_Angelil14}, and the resulting nucleation rates and bubble properties were found to differ somewhat from classical nucleation theory.
In this work we simulate a similar liquid, just at a lower level of superheating to avoid homogeneous nucleation, and we add model heat spikes to probe their ability to induce bubble formation. The simulations 
presented here provide an atomistic description of heat-spike-induced bubble nucleation.
\end{section}
%
\begin{section}{Heat spike model}
The classic heat spike model underlies the theory of the bubble detector's functionality and induced bubble nucleation, and was developed by Seitz in 1958 \cite{Seitz_paper}.
It has since been used to calibrate various detectors. However, its assumptions and predictions on the nanoscale process of bubble nucleation remain untested.
The model describes the formation of bubbles in highly localized and hot regions.
The formation process of a macroscopic bubble can be split into two parts:
the formation of a microscopic protobubble or nucleus, followed by
the growth phase into a macroscopic, gas-filled bubble.
Because we wish to determine the deposited energy and track length thresholds for stable bubble formation, in this paper our focus remains on the first part.

A spherical bubble in mechanical equilibrium experiences a few competing influences: the outwards pressure of the vapor, the pressure from the surface tension acting to collapse the bubble, and the fluid pressure pushing inwards.
This leads to the condition
\begin{equation}
 R_{c} = \frac{2\gamma}{\Delta p} \; ,
 \label{eq:Rcrit}
\end{equation}
where $R_{c}$ is the minimal critical radius, $\gamma$ the surface tension, and $\Delta p = p_{v} - p_{l}$ the pressure difference
between the (low) pressure $p_{l}$ in the superheated liquid and the vapor pressure inside the bubble $p_{v}$.

The energy required for the formation of a bubble with critical radius $R_{c}$ is the sum of surface energy and the energy of vaporization
\begin{equation}
 E_{m} = 4\pi\gamma R_{c}^{2} + \frac{4\pi}{3}R_{c}^{3}n_{v}H_{s}  , 
 \label{eq:Ecrit}
\end{equation}
where $n_{v}$ is the number of moles per unit volume at the equilibrium pressure and
$H_{s}$ is the heat of sublimation per mole. This means that bubbles are formed if sufficient energy is deposited along a track of critical length $l_{c} = \Lambda R_{c}$ or shorter.
The classical heat spike model assumes that the critical length is equal to the diameter of the critical bubble, i.e., $\Lambda = 2$.
The deposited energy $E_{dep}$ is calculated using the mean energy deposited per unit distance, or linear energy transfer (hereafter LET):
\begin{equation}
 LET_{c} = \frac{E_{m}}{\Lambda R_{c}} \leq \frac{dE_{dep}}{dx}.
 \label{eq:LET}
\end{equation}
\end{section}
%
\begin{section}{Simulation setup}
We use the classical molecular dynamics code, the Large-scale Atomic/Molecular Massively Parallel Simulator (or LAMMPS) \cite{LAMMPS}. 
We use simulation boxes of length $L=323.6\,\sigma$
with periodic boundary conditions.
The interaction between the particles is described by a truncated force-shifted Lennard-Jones (LJ) potential:
\begin{equation}
    u_{TSF}(r) = \left\{ \begin{array}{l l}
    &u_{LJ}(r) - (r-r_{c}) u'_{LJ}(r_{c}) - u_{LJ}(r_{c}) \\
    &0 \hspace{3.75cm} \textit{if} \quad r > r_{c},
    \end{array} \right.
    \label{eq:TSFLJpotential}
  \end{equation}
with the well-known Lennard-Jones potential $u_{LJ}$, and a cutoff distance $r_{c}=2.5\,\sigma$, where the potential and force smoothly approach zero.
At our run temperature ($T=0.855\,\epsilon/k_{B}$) the truncated force-shifted LJ fluid has a surface tension of $\gamma=0.0895\,\epsilon\sigma^{-2}$, a latent heat of $H_{s}=6.9\,\epsilon$, and an equilibrium pressure of $P_{eq}=0.0461\,\epsilon\sigma^{-3}$ ($\simeq p_v$; see \cite{Diemand14_Angelil14, Angelil14} for details).
At the simulation density ($\rho=0.5792\,m\sigma^{-3}$) the liquid pressure is $p_{l} = 0.028\,\epsilon\sigma^{-3}$, which results in a minimal critical bubble radius of $R_{c}=10.47\,\sigma$ and required energy deposition of $E_{m}=2164\,\epsilon$, where $\sim2041\,\epsilon$ is used for vaporization and ​$\sim123\,\epsilon$ goes into work against ​surface​ ​tension.

LJ fluids are often used to model noble gases.
Fortunately our LJ fluid also has thermodynamic properties similar to those of the liquids used in dark matter search experiments.
For example C$_{2}$ClF$_{5}$ used in the SIMPLE experiment can be approximated quite well using the following factors
when converting from LJ to SI units: $\sigma=0.501$ nm, $\epsilon=377.65\,k_{B}$K = 0.0325 eV and $m= 4.466\,$ g/mol. At 9\textcelsius  (the running temperature of SIMPLE) this results in an equilibrium pressure and a surface tension which both agree to within
26 percent between C$_{2}$ClF$_{5}$ \cite{Okada} and our LJ fluid.
The main difference is that our simulations have a higher amount of superheat, a smaller critical radius ($10.47 \sigma = 5.26$ nm vs. around 40 nm in SIMPLE \cite{SIMPLE3}) and a much smaller required
energy deposition ($E_{m}=2164\,\epsilon = 70.42\,eV$ vs. a few keV in SIMPLE).
Simulating heat spikes at comparable thermodynamic conditions to those in the experiments is therefore possible with the same interaction potential,
but will require significantly larger simulation volumes (which is computationally very expensive) to accommodate the formation of large critical bubbles at constant ambient pressure.

To prepare the liquid in a superheated, metastable state, we use a procedure similar to that of Diemand \textit{et al.} \cite{Diemand14_Angelil14}:
First 19'652'000 atoms are placed on a lattice corresponding to a density of $\rho = 0.58\,m\sigma^{-3}$.
The particles are then randomly assigned velocities corresponding to 
a temperature of $T=0.95\,\epsilon/k_{B}$. 
A fixed time step of $\Delta t=0.0025\,\tau=0.0025\sqrt{\sigma^{2}m/\epsilon}$ was used throughout all runs.
A run of 10'000 time steps is initiated under \textit{NVT} integration
(constant particle number, volume, and temperature ensemble) at a temperature of $T = 0.95\,\epsilon/k_{B}$.
These are stable liquid conditions, used to equilibrate the atoms which were initially placed on a lattice.
The liquid is superheated by lowering the temperature from $T = 0.95\,\epsilon/k_{B}$ to $T = 0.855\,\epsilon/k_{B}$ over a relatively short period of 15'000 steps as an \textit{NVT} ensemble.

Another \textit{NVT} run of 15'000 steps is performed to stabilize the system at the new temperature $T = 0.855\,\epsilon/k_{B}$.
The thermostat is then turned off, and the simulation allowed to continue for 450'000 time steps with simple direct integration of the classical equations of motion (i.e., as a microcanonical or \textit{NVE} ensemble).
A typical simulation spreads 256 MPI tasks over the same number of cores, and runs for between six and eight hours.

Bubbles are identified with the same method as in \cite{Watanabe,Diemand13_Angelil13}: The simulation volume is divided up into cubic cells of size $(3\sigma)^3$.
Cells with a density $\rho < 0.2\,m\sigma^{-3}$ are marked as vaporized regions and are recursively linked with neighboring vapor cells into a connected vapor bubble.
As expected, these runs did not result in spontaneously nucleated bubble formation due to the fact that the density was too close to the stable equilibrium value, making the free energy barrier between the liquid and the gaseous phase too high for spontaneous homogenous bubble nucleation.

Following Seitz's model, we mimic the nuclear recoil energy injection heat spike by assigning higher velocities to all particles within a cylindrical region (see the upper panels of Fig. \ref{fig:snapshots}).
Each simulation continues for 200'000 to 500'000 time steps after the heat spike, in order to determine whether or not a stable, growing bubble manages to form out of the heat spike.
In total 40 simulations with different deposited energies and track lengths were performed.

\begin{figure}[ht]
  \includegraphics[width=0.15\textwidth]{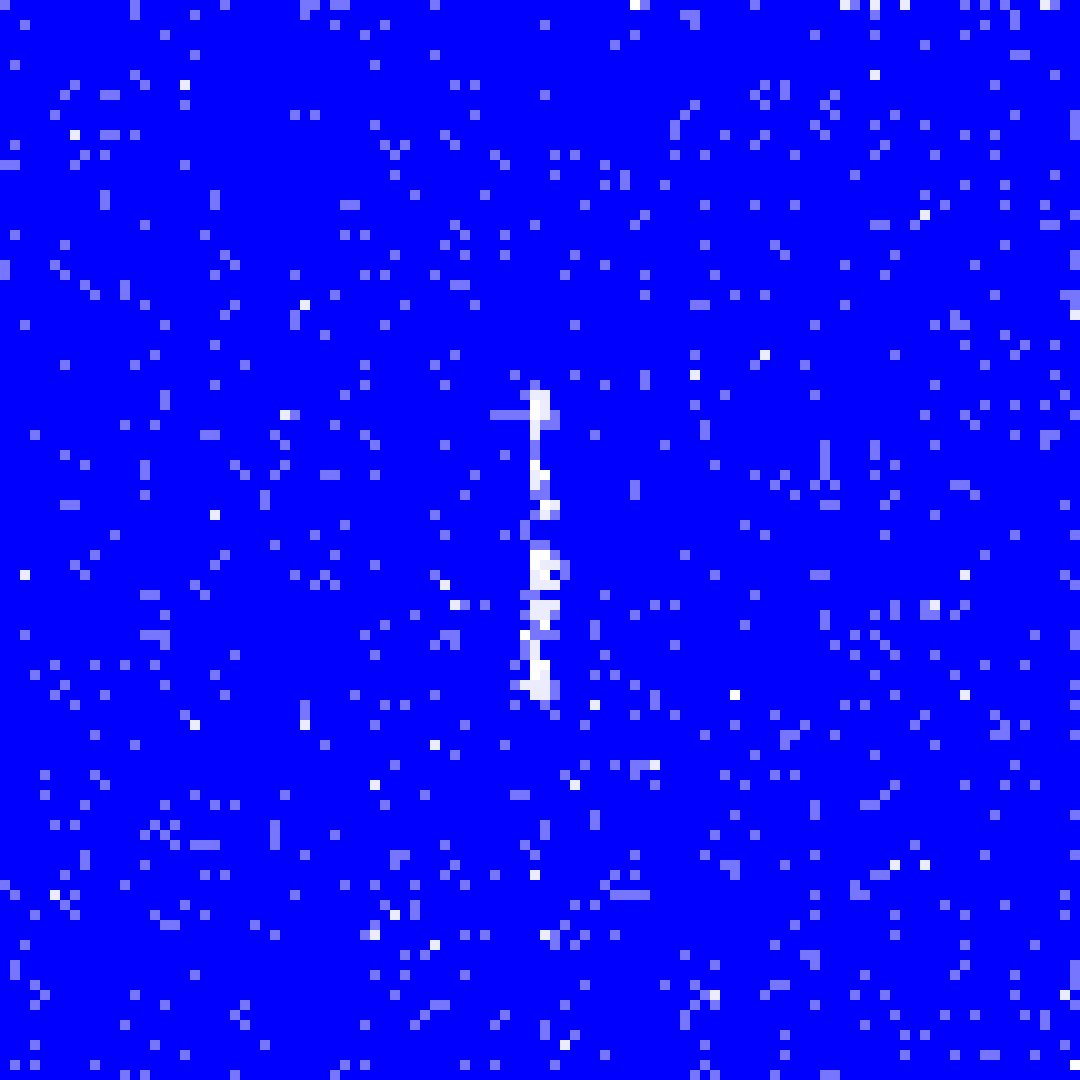}\hspace{3pt}\includegraphics[width=0.15\textwidth]{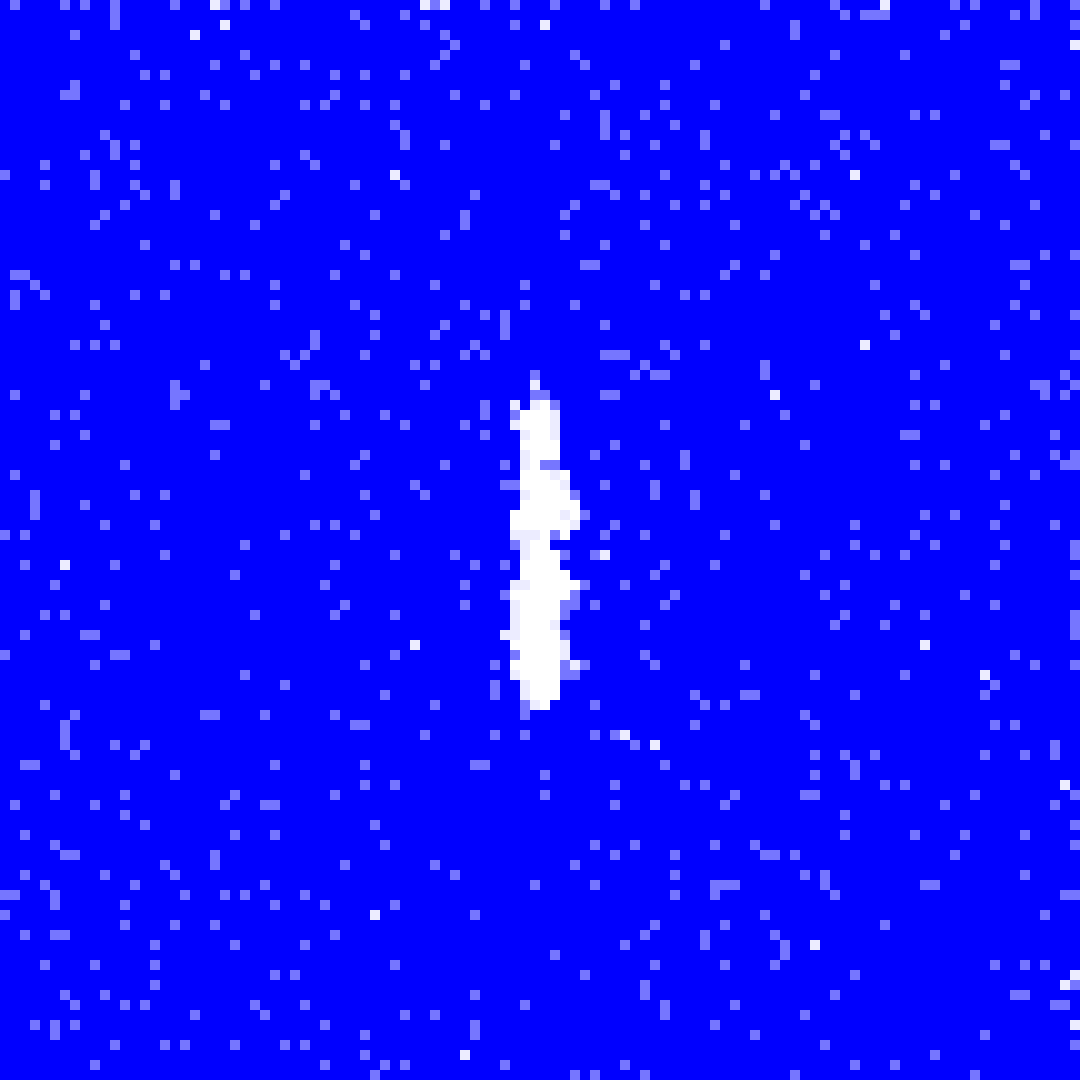}\hspace{3pt}\includegraphics[width=0.15\textwidth]{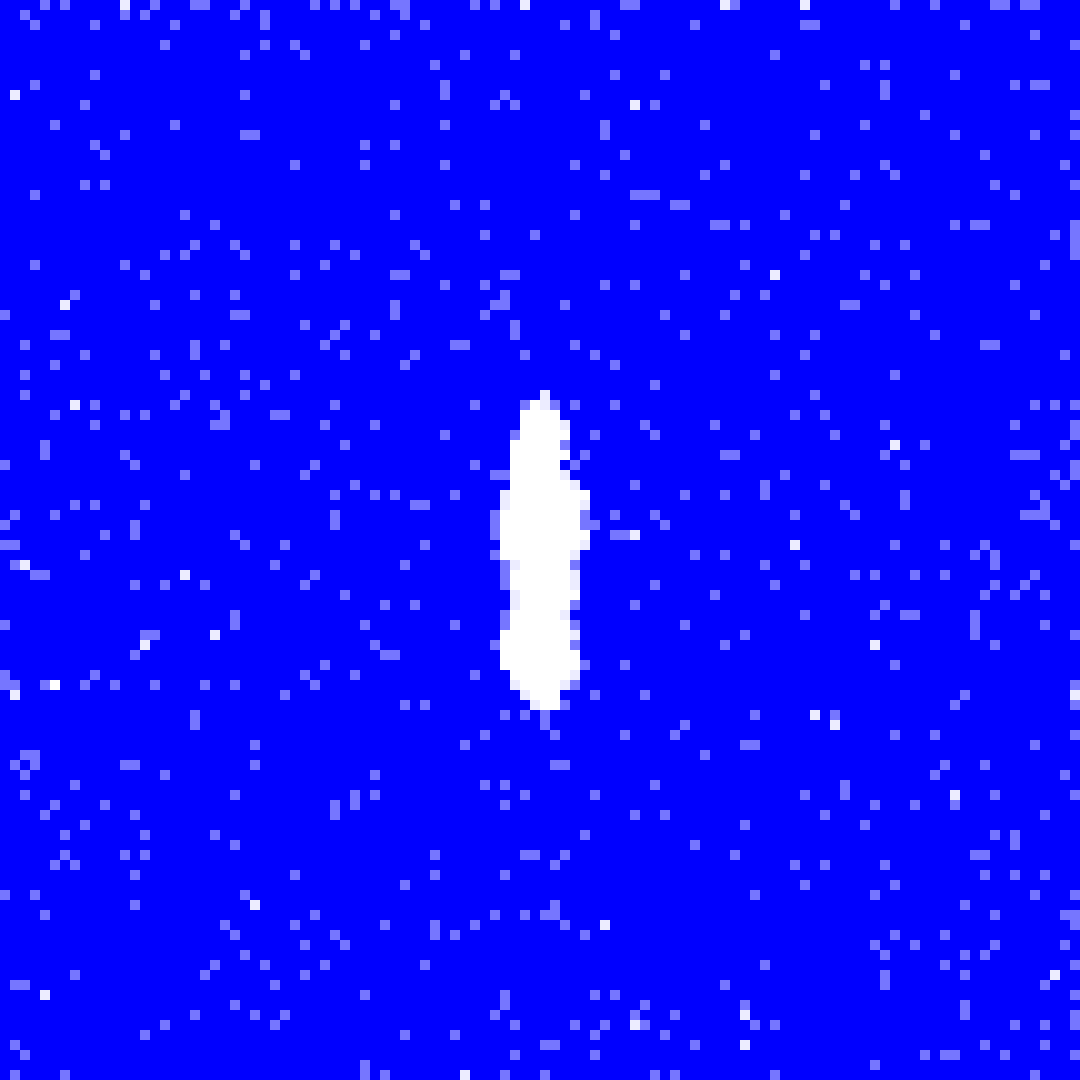}\vspace{3pt}
  \includegraphics[width=0.15\textwidth]{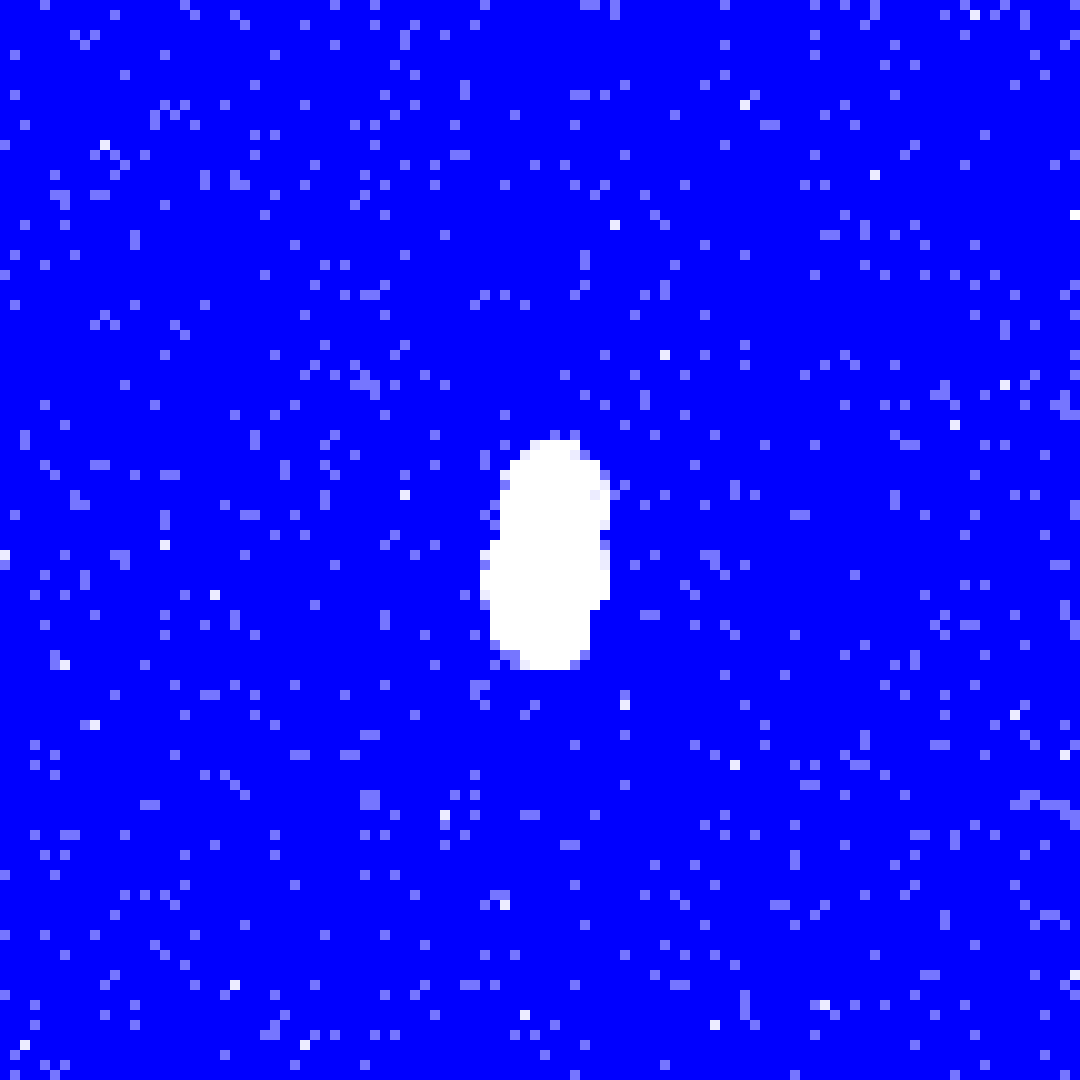}\hspace{3pt}\includegraphics[width=0.15\textwidth]{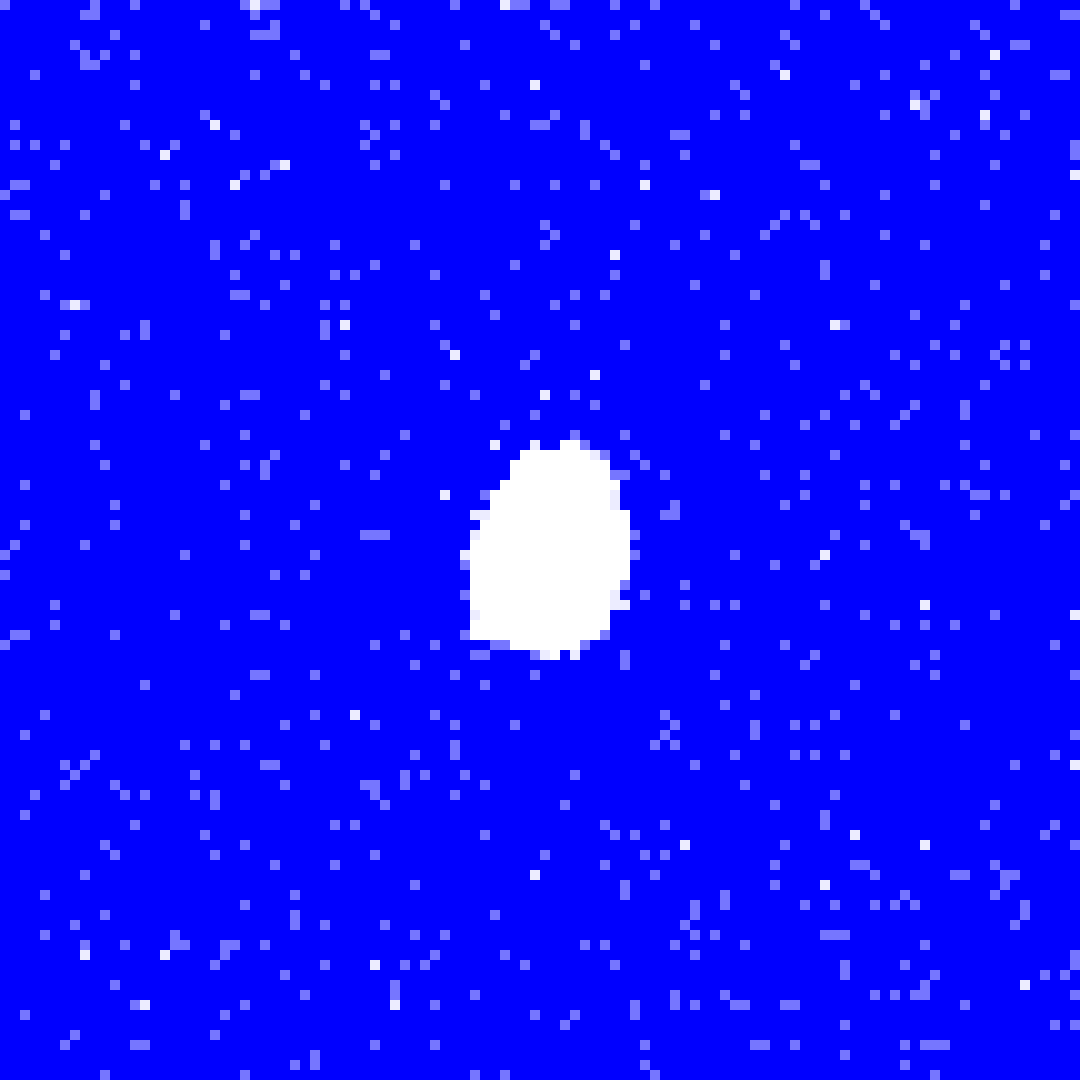}\hspace{3pt}\includegraphics[width=0.15\textwidth]{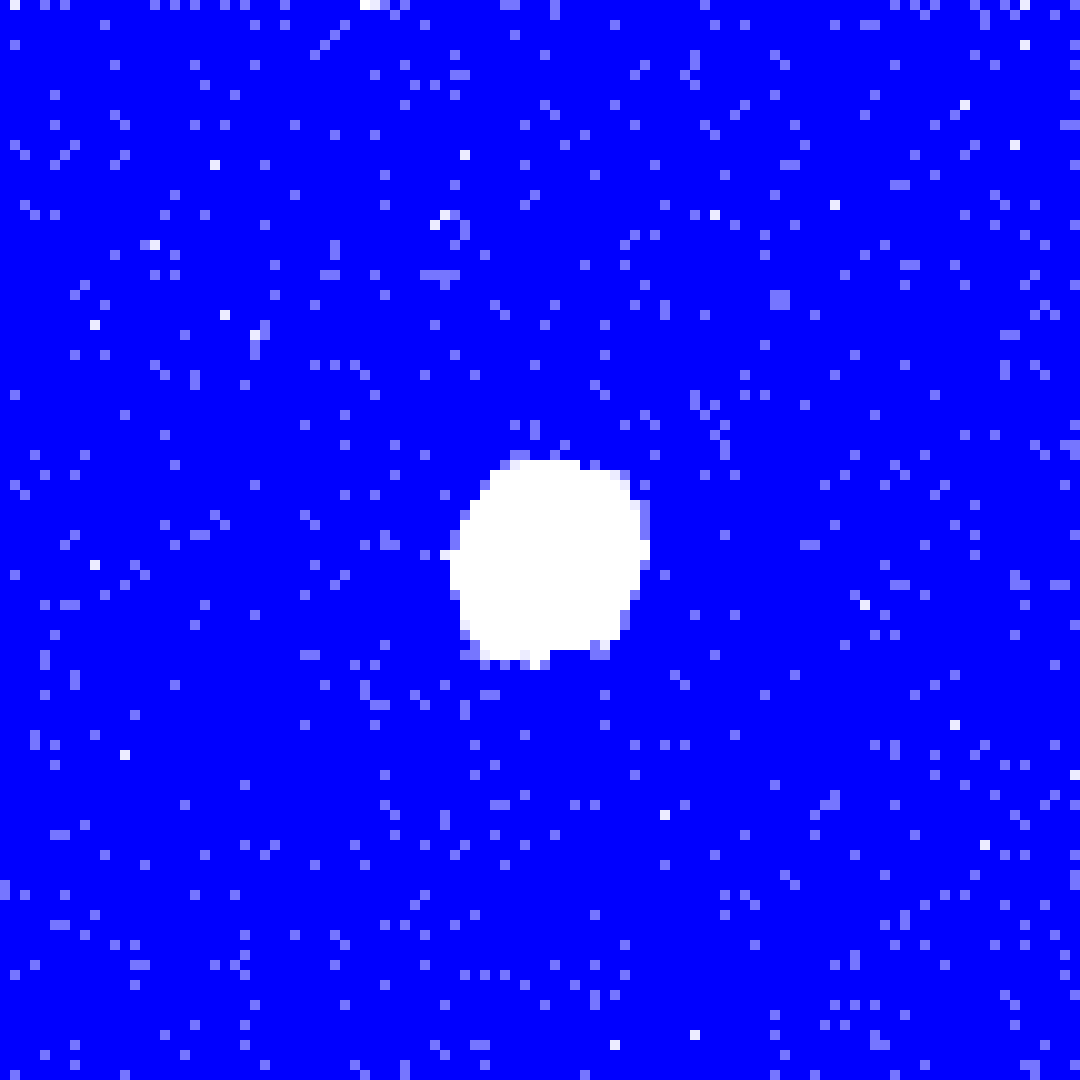}\vspace{3pt}
  \caption{(Color online) Projected snapshots of run at $7.5\,\tau$, $37.5\,\tau$, $125\,\tau$, $500\,\tau$, $812.5\,\tau$, and $1250\,\tau$ after a heat spike of $12k\,\epsilon$ along a length of $8\,R_{c}$.}
  \label{fig:snapshots}
\end{figure}
\end{section}

%
\begin{section}{Simulation results}
The bubble nucleation event can be split into three stages (see Fig. \ref{fig:growth}).
\begin{enumerate}
 \item The first stage is preceded by a heat spike, which results in an explosion of the bubble radius \footnote{We assume spherical bubbles when converting the observed bubble volume to a radius. However, immediately after the heat spike, the bubbles are far from spherical (see Fig. \ref{fig:snapshots}). Large, stable bubbles eventually do become spherical \cite{Angelil14}}, with an initial expansion rate of about $0.40\,\sigma$/$\tau$ and up to $1.0\,\sigma$/$\tau$ for simulations with large energy deposition.
In all cases we observe an overshoot right after this early explosive phase and the bubble volume decreases slightly.
If the heat spike produces a bubble with a radius of about $10\,\sigma$ to $15\,\sigma$ at the end of the first stage, it generally is stable and will transition into the second growth stage.
Bubbles which are too small, however, are unstable, and recollapse completely.
 \item During the second stage we observe linear growth perturbed by small oscillation patterns.
Since the frequency of the oscillations matches the speed of sound inside the medium over the length of the box, the patterns can be explained by a pressure wave propagating through the simulation box.
 \item In the third stage, the bubble grows linearly until the system pressure increases due to the finite simulation box size, which unrealistically slows the growth of large bubbles in the final stages of our simulations.
Typical average expansion speeds over the linear regime are around $0.0085\,\sigma/\tau$ to $0.0130\,\sigma/\tau$.
These growth rates are in good agreement with Ang\'elil \textit{et al.} \cite{Angelil14}, where the bubble growth of homogeneous nucleation was investigated under similar conditions.
\end{enumerate}

We tested the stability of our simulations by observing the box pressure after the heat spike throughout all the runs and found a stable pressure
of $P_{init} = 0.0288 \pm 0.002 \,\epsilon\sigma^{-3}$ averaged over 10'000 time steps.
This tells us that the simulation box size and number of atoms that we use to model our fluid, are chosen reasonably and do not effect the physics of the problem investigated.

\begin{figure}[ht]
 \includegraphics[width=0.5\textwidth]{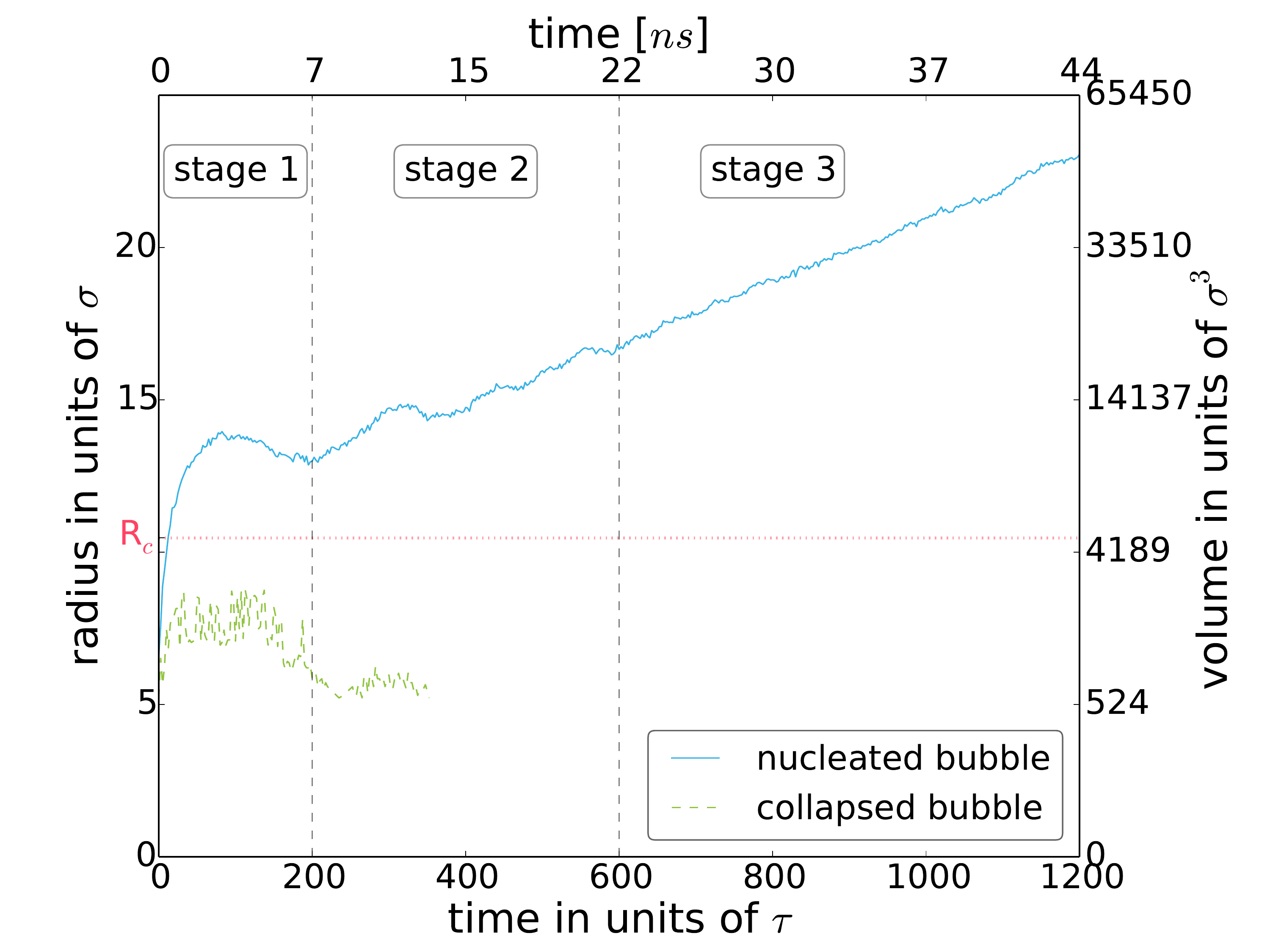}
 \caption{(Color online) Three-stage bubble growth plot of a run with a heat spike of $5000\,\epsilon$ (or $3000\,\epsilon$) along a track of $1\,R_{c}$ yielding a stable growing (collapsing) bubble.}
 \label{fig:growth}
\end{figure}

Figures \ref{fig:E} and \ref{fig:LET} show the outcome of simulations with a fixed cylinder radius of $2\,\sigma$ in a plot of the deposited energy $\Delta E$ relative to the length of the cylinder $l_{c}$.

\end{section}
%
\begin{section}{Comparison to theory} 

Seitz's model predicts that the energy threshold $\Delta E$ in our simulated fluid is $2164\, \epsilon = 70.3\, eV$. From the energy inputs required to generate a growing bubble (Fig. \ref{fig:E}) we find $\Delta E (\Lambda=2) = 4166.1 \pm 125.0\, \epsilon = 135.4 \pm 4.1\, eV$, which is about a factor of 2 higher.
The length scale over which the energy must be deposited ($\Lambda R_{c}$) is also underestimated by the theory:
According to Seitz, the energy contributing to one bubble formation event must be confined within a length of $2\,R_{c}$; i.e., his model assumes $\Lambda=2$.
The simulations however show that the critical energy can be spread over up to 4 critical lengths without losing bubble formation efficiency\textemdash a discrepancy of a factor of 2.
On tracks longer than $\Lambda=4$, the required total energy for bubble formation is higher, since only a fraction of this now too widely spread energy contributes to a single bubble formation event.

\begin{figure}[ht]
  \includegraphics[width=0.5\textwidth]{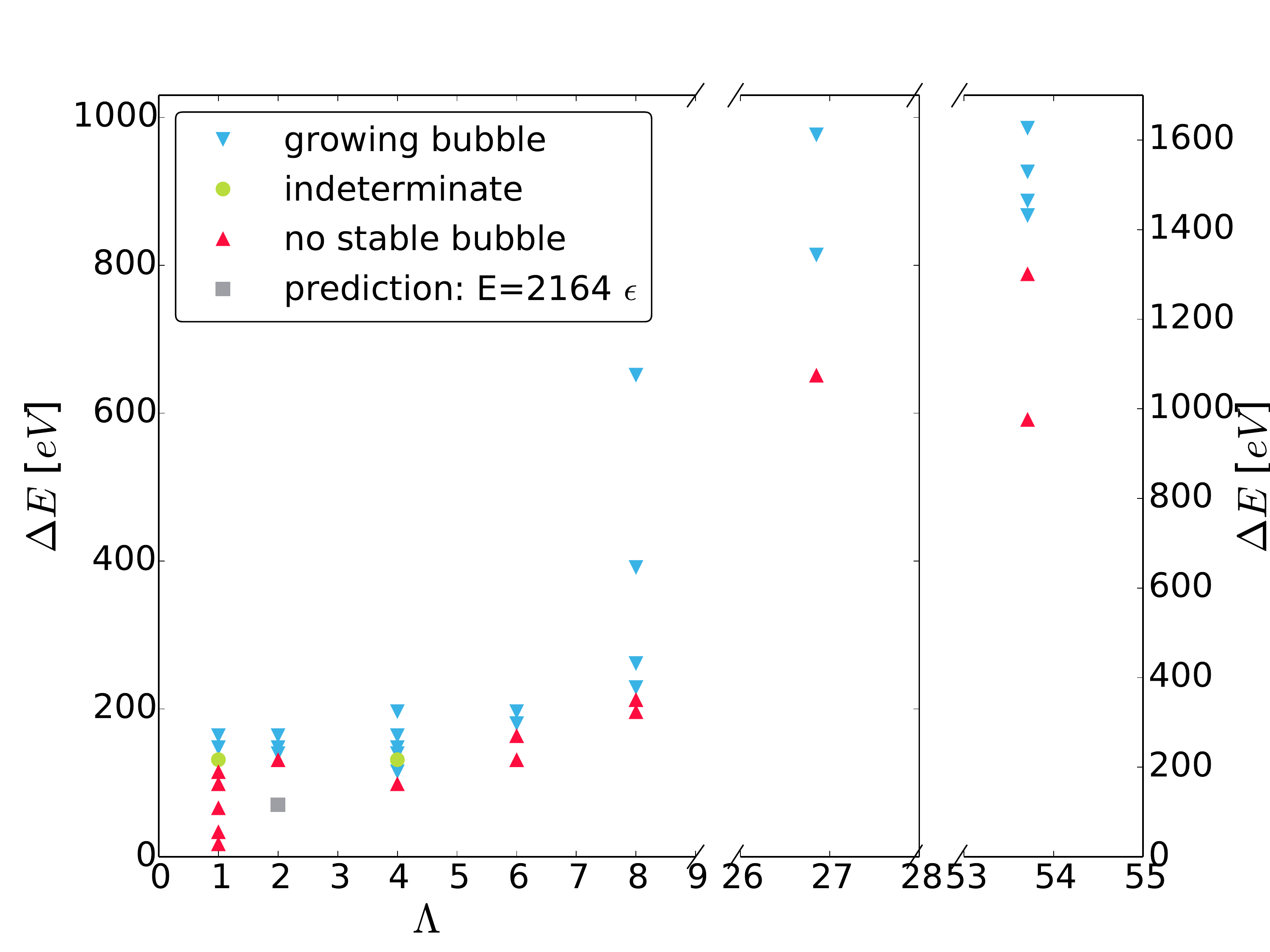}
  \caption{(Color online) Energy plot of all runs with cylinder radius $r_{cyl}=2\sigma$ and cylinder length $l_{cyl}=\Lambda R_{c}$.
  The critical energy to nucleate a bubble lies around $4500\,\epsilon$, and can be spread over up to a length of four critical radii, which is in both cases a factor of 2 greater than the theoretical predictions.}
  \label{fig:E}
\end{figure}

For analysis it is also convenient to consider the relationship between the LET [see Eq. (\ref{eq:LET})] and the cylinder length (see Fig. \ref{fig:LET}).
For $\Lambda = 4$ and all larger values we find LET values which are consistent with a constant value of $69.31 \pm 4.41\, \epsilon/\sigma = 4.50 \pm 0.29\, keV/\mu m$. The theoretical prediction for the LET is quite accurate, just somewhat larger: $89.80\, \epsilon/\sigma = 5.83\, keV/\mu m$.

The LET plots show that the Seitz predictions for the LET are accurate, provided that the LET is sustained for at least four times the critical radius.
Twice the critical radius is insufficient. Even up to lengths of $\Lambda = 26.88$ and $\Lambda = 53.71$ for $\alpha$-particle runs, the results match the predictions.

\begin{figure}[ht]
  \includegraphics[width=0.5\textwidth]{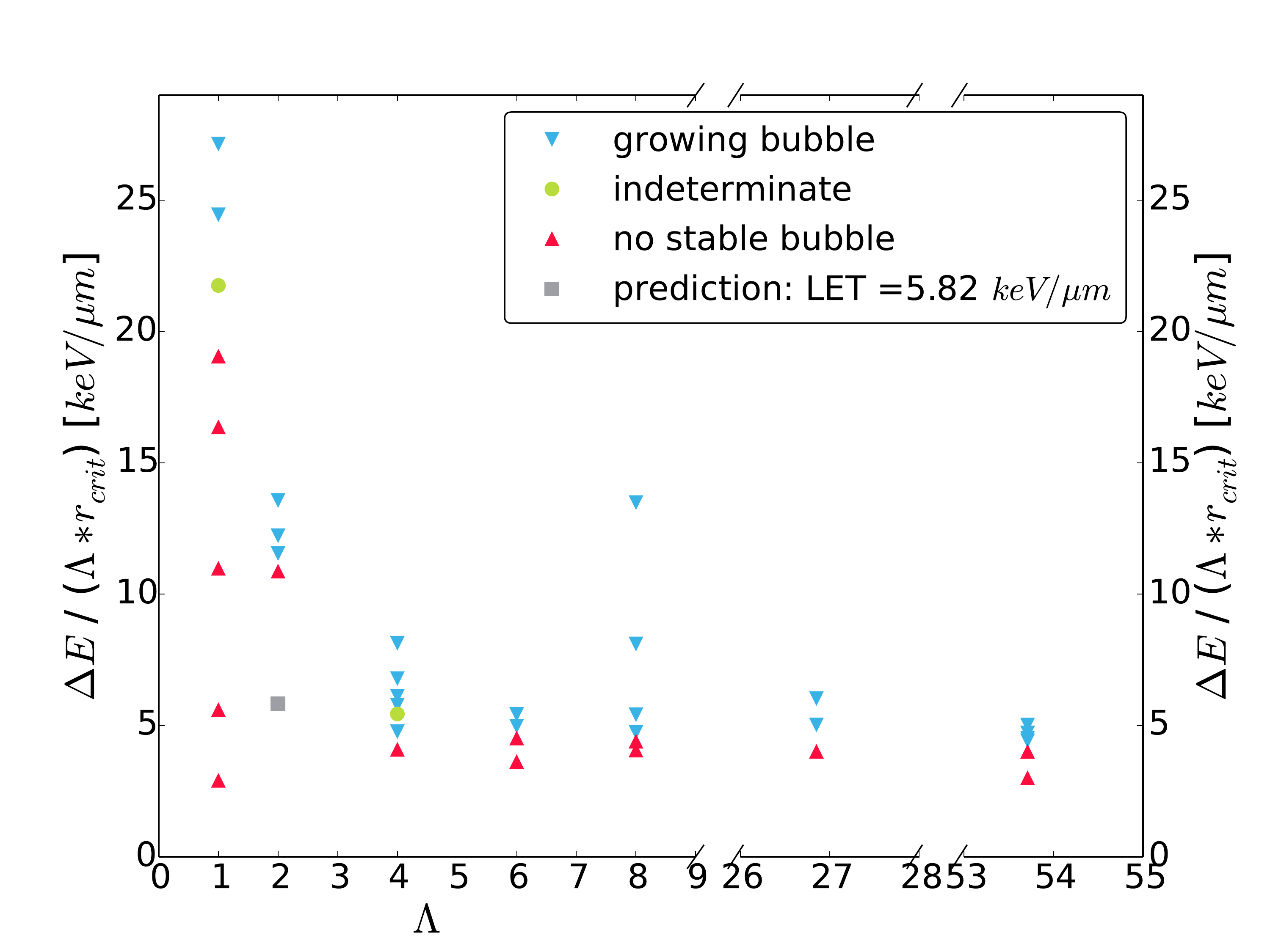}
  \caption{(Color online) LET plot of same runs as in Fig. \ref{fig:E}.
  The LET matches the theoretical prediction, provided a length of at least four critical radii.}
  \label{fig:LET}
\end{figure}

We have also investigated the dependence of the deposited energy on the cylinder radius.
For small cylinder radii ($0.744\,\sigma$ and $2\,\sigma$) and small $\Lambda$, the required energies are almost equal, even though the number of atoms in the cylinder differs by a factor of $\left(\frac{2\sigma}{0.744\sigma}\right)^{2} \approx 7.22$. 
For much higher cylinder radii, 
the required energy must be higher also, because heat diffusion is accelerated with more atoms.

Although the model makes some quite accurate predictions (e.g. for the LET), it contains several unrealistic simplifying assumptions:
\begin{itemize}
\item The microbubble thermophysical properties are assumed to be the same as in a planar, static, isothermal interface in equilibrium between bulk liquid and bulk vapor. 
The inclusion of a more realistic surface tension, specifically, its dependence on bubble size \cite{tolman1949,kyokotanaka2015}, should be addressed in future refinement of the model.
\item Viscosity is not included in the model.
\item Microbubbles are assumed to be spherical and to have a steplike vapor-liquid transition region. But critical bubbles are nonspherical and the width of their transition regions can be comparable to their radii \cite{Angelil14}.
\item A more realistic model would include the complicated and evolving non-isothermal bubble temperature profiles due to the latent heat of transformation as well as compressive heat in the fluid directly outside the bubble due to rapid bubble expansion. Significant non-isothermal effects are found in our heat spike simulations and also in homogeneous bubble nucleation simulations; see \cite{Angelil14}.
\end{itemize}

\end{section}
%
\begin{section}{Dissipation of heat}
Seitz's model assumes that a critically-sized bubble forms before the deposited energy diffuses out of the critical volume. As we shall show, this is not the case for our simulated liquid. To establish a criterion related to the dissipation of the deposited energy, we assume that the energy is
deposited uniformly in a spherical region of radius $R_{0}$ smaller than $R_{c}$. The volume shall be equal to a cylinder volume of length $2 R_{c}$ and cross section $\pi a^{2}$, similar to a track covered by a recoil. $a$ is chosen such that $\frac{4\pi}{3}a^{3}$ is the mean molecular volume, where $a$ is the van der Waals radius. Then, the critical time $\tau_{c}$ for the dissipation of heat in a spherical spike of radius $r$ is determined by setting $r=R_{c}$:
\begin{equation}
 \tau_{c} = \frac{R_{c}^{2}}{4D},
 \label{eq:crittime}
\end{equation}
in which $D$ is the diffusion coefficient for heat. This is an approximation, yet all the quantities are mostly constant with variable pressure, so the values should be accurate to a factor of 2. With Eq. (\ref{eq:crittime}) one can assign an average minimum velocity with which the walls of a bubble must expand to achieve stability before the deposited heat is dissipated,
\begin{equation}
 v_{c}=\frac{R_{c}}{\tau_{c}}
 \label{eq:vcrit}
\end{equation}
It is interesting to compare this velocity with the speed of sound in the liquid $v_{sound}$.
Comparing the literature values, Seitz finds that $v_{c}$ is almost a factor of 10 less than $v_{sound}$.
Consequently, the average velocity with which the walls of the bubble have to expand in order to form a bubble during a time around $\tau_{c}$ is subsonic (see \cite{Seitz_paper} for more details).
The kinetic energy inside a sphere at the critical radius is expected to decay exponentially. By fitting an exponential decay function to the simulation data, the diffusion time and velocity can be obtained: $\tau_{c}=3.41\,\tau=6.82\times10^{-12}\,s$ and $v_{c}=3.54\,\sigma/\tau = 601.03\,m/s$ (see Fig. $\ref{fig:Ekindens}$).
These values can be compared to the diffusion time calculated with literature values for Lennard-Jones MD simulations, according to Eq. (\ref{eq:crittime}): $v_{diss}=96.02\,m/s$.
This value is in agreement with the simulations if one takes into account that the analysis assumes perfectly spherical heat spike bubbles, which is not the case in our runs (see Fig. \ref{fig:snapshots}).
Seitz assumes that critical bubble formation occurs within the dissipation time scale $\tau_{c}$.
We find that this is clearly not the case:
The kinetic energy to density comparison makes it clear that the heat dissipation is significantly faster than the bubble expansion: the average kinetic energy in a sphere of critical radius disappears before the bubble number density drops.
The measured initial expansion velocities imply that the heat dissipation is a factor of $3-10$ times faster then the bubble expansion.

On the other hand, Seitz compares his diffusion velocities with acoustic velocities of the same media and concludes that the diffusion is subsonic. The contrary is the case for our simulations.
The expansion velocities of simulated bubbles are 3 to 10 times slower and lie in the same range as the speed of sound for a monoatomic gas.

In Fig. \ref{fig:Ekindens}, we show the evolution of density and average of the kinetic energy, which is proportional to temperature, averaged over a fixed volume of radius R$_{c}$ centered on the midpoint of the heat spike, which was inserted at time t$=0$.
After a time of about 60 $\tau$ the density values do not change significantly anymore inside this volume, since the bubble exceeds the critical size after this time.
A quick dimensional analysis with values from \cite{koreanbull} lets us expect a thermal conductivity of 0.1 $J K^{-1} m^{-1} s^{-1}$, and we get a heat flow of $\frac{Q}{t} = 3.4 \,J/\tau_{c}$.
The internal energy per atom is $U = \frac{3}{2}k_{B}T \approx 10^{-21} J$ within a time of $\tau_{c}$, which means that an energy of about 32'000 atoms can be transported out of the critical volume.
These dimensions lie well within the range of our simulation results.

It would also be very interesting to measure the evolution of pressure inside the bubble volume after the heat spike, which then would give insights to the surface tension at the bubble boundary, i.e., the surface tension as a function of the bubble radius. Unfortunately, the measurements were much too noisy to deduce anything from them.

\begin{figure}[ht]
  \includegraphics[width=0.5\textwidth]{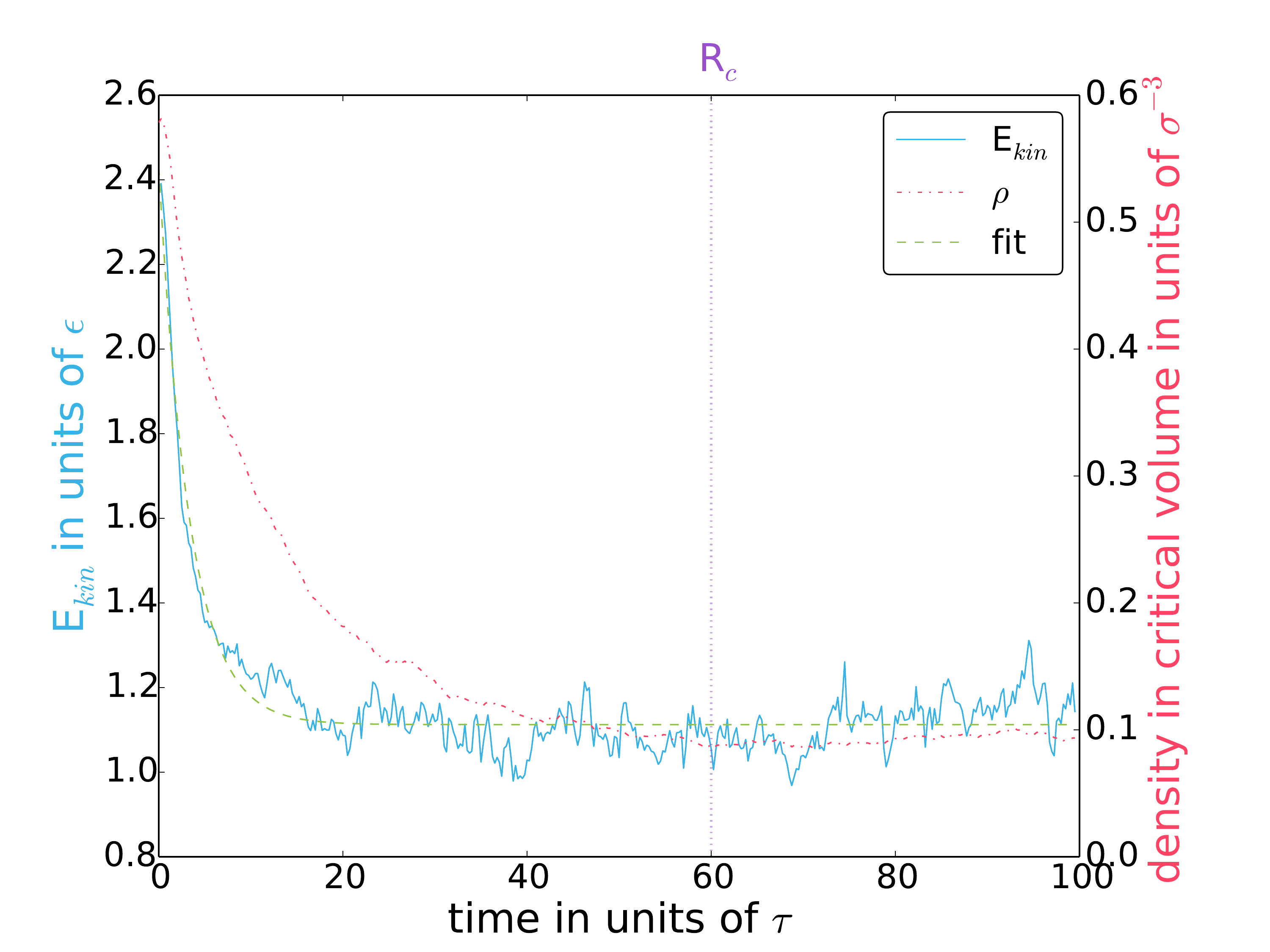}
  \caption{(Color online) Evolution of kinetic energy and density within a spherical volume of radius $R_{c}$. By fitting the data for the kinetic energy with an exponential decay function $E_{kin}(t) = A\cdot \exp(K\cdot t) + b$ the relaxation time $\tau_{c} = K^{-1}$ can be obtained. The results are (A, K, b)$=$(1.3195, 7.4640e-4, 1.1124) with squared diagonal elements of the covariance matrix (6.8294e-7, 3.7630e-19, 8.5889e-11) in units of ($\epsilon$, $0.0025\cdot\tau$, $\epsilon$).}
  \label{fig:Ekindens}
\end{figure}

\end{section}
%
\begin{section}{$\alpha$-particle discrimination}
Superheated liquids function as threshold detectors.
All particles able to deposit more energy than this (tunable) detector threshold will lead to bubble nucleation and an observable acoustic signal.
A setup to search for WIMPs is also sensitive to $\alpha$ particles, because they are able to deposit higher energies.
Tracks from $\alpha$ particles are far longer, and their discrimination is essential for a convincing dark matter detection.
Empirically it has been found that the longer $\alpha$-particle tracks lead to larger acoustic signals compared to neutrons \cite{alpha_discr,COUPP2} (which have tracks similar to typical WIMPs), presumably because a large number of bubbles can form along a longer track.
However the origin of the louder signal remains unclear.
Currently there is some discussion around robustness of the $\alpha$-particle discrimination\cite{alpha_discr,SIMPLE3}.
For this reason, we go beyond testing Seitz's model, and simulate far longer tracks (see Fig. (\ref{fig:E}) and (\ref{fig:LET})).
Only a small fraction of an $\alpha$ track can be simulated, because the computation cost of simulation boxes of lengths in the $\mu$m range would be too large.
For the $\alpha$ runs we set the track length equal to the box size $L$. We also ran additional large box $\alpha$ simulations 
with $L=647.1\,\sigma$ and 157'216'000 atoms and found consistent results:
 Fig. (\ref{fig:LET}) shows that even for long tracks with $\Lambda$ = 53.71, the required LET is the same.
$\alpha$ tracks above the required LET produce several stable bubbles separated by only 4 to 5 critical radii.
Rescaled to an experiment such as SIMPLE ($R_{c}\sim40$ nm), this indicates a bubble density of 5 per micrometer and around 200 microbubbles on a typical $\alpha$-particle track.
The acoustic signal is generated mostly in the linear growth phase \cite{alpha_discr} and a large number of linearly growing (and later merging) bubbles explains the louder signal compared to the single bubble case.
\end{section}
%
\begin{section}{Conclusions}
We have studied dark matter interaction induced bubble nucleation using direct molecular dynamics simulations of the process. Our molecular dynamics simulations of dark matter bubble detectors are the first attempt of an atomistic, computational description of the heat-spike-induced bubble nucleation event. Our results qualitatively confirm the general framework assumed in the classical heat spike model by Seitz, but also show some interesting quantitative differences. 

\begin{itemize}
 \item In comparison to the model predictions, for stable bubbles to be successfully nucleated, we find that the deposited energy must be approximately twice as much, whereas the length over which it is deposited can be twice as long. This means that the model predictions actually are rather accurate for the linear energy transfer prediction for tracks, which are at least 4 critical radii long, twice the length assumed in the model. The measured heat diffusion time scale is shorter than the time it takes to form a critical size bubble, contrary to what is assumed in the model, which could explain the larger-than-expected required energy deposition for bubble formation. 

 \item In the bubble-chamber dark matter particle search community there is still some discussion on the discrimination of WIMP signals from the $\alpha$-particle background \cite{alpha_discr}, which depends on exact energy predictions from the model of induced bubble nucleation.
By giving corrections to the energy predictions of the model, this paper and further work could help in the correct calibration of the detectors and analysis of data, especially the discrimination of $\alpha$-particle background, and generally gives a better understanding of the measuring processes.
\end{itemize}

Large future simulations would be needed to explore more stable fluid critical bubbles as large as in the detectors. The rescaled thermodynamic properties of our LJ fluid are comparable to some of the detector fluids; however to closely match one specific detector it would be worthwhile to simulate more specific and more complicated fluids. Our simulations demonstrate complications (bubble properties, non-isothermal effects, etc.) which are neglected in the classical heat spike model by Seitz, which might provide an opportunity for constructing more realistic models of the process. 
\newline

\end{section}

%

\begin{acknowledgments}

 We thank Kyoko and Hidekazu Tanaka, Tom Girard, and the SIMPLE team for valuable discussions and feedback. We also thank the anonymous referees for useful comments and suggestions. J.D. and R.A. acknowledge support from the Swiss National Science Foundation.

 \end{acknowledgments}

 \bibliography{heat_spike_paperJDRAPD}

\end{document}